# Thickness dependence of electronic and crystal structures in VO$_2$ ultrathin films: suppression of the collaborative Mott–Peierls transition

D. Shiga[1,2], B. E. Yang[1], N. Hasegawa[1], T. Kanda[1], R. Tokunaga[1], K. Yoshimatsu[1], R. Yukawa[2], M. Kitamura[2], K. Horiba[2], and H. Kumigashira[1,2,*]

[1] *Institute of Multidisciplinary Research for Advanced Materials (IMRAM), Tohoku University, Sendai, 980–8577, Japan*

[2] *Photon Factory, Institute of Materials Structure Science, High Energy Accelerator Research Organization (KEK), Tsukuba, 305–0801, Japan*

**Abstract**

Through *in situ* photoemission spectroscopy, we investigated the change in the electronic and crystal structures of dimensionality-controlled VO$_2$ films coherently grown on TiO$_2$(001) substrates. In the nanostructured films, the balance between the instabilities of a bandlike Peierls transition and a Mott transition is controlled as a function of thickness. The characteristic spectral change associated with temperature-driven metal–insulator transition in VO$_2$ thick films holds down to 1.5 nm (roughly corresponding to five V atoms along the [001] direction), whereas VO$_2$ films of less than 1.0 nm exhibit insulating nature without V-V dimerization. These results suggest that the delicate balance between a Mott instability and a bandlike Peierls instability is modulated at a scale of a few nanometers by the dimensional crossover effects and confinement effects, which consequently induce the complicated electronic phase diagram of ultrathin VO$_2$ films.

[*]Author to whom correspondence should be addressed: kumigashira@tohoku.ac.jp

# I. INTRODUCTION

The metal–insulator transition (MIT) of vanadium dioxide ($VO_2$) [1], which is one of the most controversially discussed phenomena for decades, is particularly intriguing because both the structural transition and electron correlation contribute to it [2–15]. The structural phase changes from high-temperature metallic rutile ($P4_2/mnm$) to low-temperature insulating monoclinic ($P2_1/c$) across the MIT near room temperature [2,3]. As a result, the MIT is accompanied by an orders-of-magnitude change in conductivity and exhibits an ultrafast response to external stimuli; thus, this phenomenon has become a central topic in modern condensed matter physics for its potential application in future electronic devices [16–22]. Further, the unusual phenomenon originating from the interplay of the electron correlation and the lattice provides an opportunity to better understand the fundamental physics of strongly correlated oxides.

In the phase transition in $VO_2$, tilting and pairing of V ions along the $c_R$ axis, which is defined as the $c$ axis of the rutile structure in the monoclinic phase, mark this structural change. Because the distances between paired V ions (0.29 nm in the high-temperature metallic rutile phase [3]) and between V ion pairs are different, the V ions in $VO_2$ are collectively dimerized along the $c_R$ axis in the insulating monoclinic phase [2,3]. Although the MIT that is concomitant with the collective dimerization of V atoms is reminiscent of the Peierls transition [6,7], the importance of strong electron correlations in $VO_2$ has also been evident for this MIT from a large number of experimental and theoretical investigations [8,9]. Therefore, the mechanism of the MIT in $VO_2$ is now mainly understood as a collaborative Mott–Peierls (or Peierls–Mott) transition [10–15]. This type of MIT has motivated researchers to clarify the role of each instability in the unusual phenomena by changing their balance via external stimuli, such as light [23–26], strain [27,28], or electrostatic injection of charge (electron doping) using field-effect transistor structures [18–22]. However, in spite of intense experimental and theoretical efforts, the origin of the MIT is still under debate.

An alternative approach to address the MIT driven by the cooperation of these two instabilities is the dimensional crossover that occurs in an artificial structure [29–35]. In the case of $VO_2(001)$ films, the "one-dimensional" V-ion chain runs along the [001] direction ($c_R$ axis), i.e., normal to the film surface, and the bandlike Peierls instability can be changed as a function of film thickness (number of V ions in the chain along the $c_R$ axis) owing to the confinement effects [29–32].

Moreover, the increment of electron-electron correlation (Mott instability) due to the dimensional crossover effects concomitant with a decrease in film thickness [33–35] may also suppress the cooperative Mott–Peierls transition in $VO_2$. Although this straightforward approach has been attempted by some groups so far [36–44], there is little consensus about the intrinsic behavior of ultrathin $VO_2$ films in reducing the film thickness because these samples suffer from significant interdiffusion of the constituent cations at the interface [36–43] as well as inhomogeneous strain from the substrate [36–38] and surface overoxidation [36–42]. Therefore, systematic control of the dimensionality while keeping the fundamental chemical and crystallographic structural parameters fixed is crucial to obtaining information on how the electronic structures and characteristic dimerization of $VO_2$ change as a function of film thickness.

Against this backdrop, in this study, we determined the electronic phase diagram of ultrathin $VO_2$ films coherently grown on $TiO_2$(001) substrates, which have chemically abrupt interfaces. By using a nonequilibrium growth process of pulsed laser deposition (PLD) that can reduce the growth temperature [45], coherent $VO_2$ films with good crystallinity were obtained while significantly suppressing the interface diffusion. Subsequently, we investigated the change in electronic and crystal structures of dimensionality-controlled $VO_2$ films via *in situ* photoemission spectroscopy (PES) and x-ray absorption spectroscopy (XAS) measurements. The PES spectra exhibited remarkable and systematic changes as a function of the $VO_2$ film thickness. (1) The characteristic spectral changes associated with the collaborative Mott–Peierls MIT remain almost unchanged down to 2.0 nm (roughly corresponding to 7 unit cells along the [001] direction). (2) In the case of film thicknesses of 1.5–2.0 nm, a pseudogap is formed at the Fermi level ($E_F$) for high-temperature metallic phase, and the spectral change across the MIT consequently weakens but still exists. (3) The pseudogap finally evolves into an energy gap when the film thickness is reduced to less than 1.5 nm, indicating the occurrence of thickness-dependent MIT at a critical thickness of 1.0–1.5 nm. In conjunction with the corresponding polarization-dependent XAS spectra, the temperature-driven MIT is accompanied by V-V dimer formation, whereas $VO_2$ films of less than 1.5 nm thickness exhibit rutile-insulating nature without V-V dimerization. These results suggest that the delicate balance between a Mott instability and a bandlike Peierls instability is modulated at a scale of a few nanometers by the dimensional crossover effects and confinement effects, which consequently induce the complicated electronic phase diagram of ultrathin $VO_2$ films.

## II. EXPERIMENT

Thickness-controlled VO$_2$ ultrathin films were grown on the (001) surface of 0.05 wt.% Nb-doped rutile-TiO$_2$ substrates in a PLD chamber connected to an *in situ* photoemission system at BL-2A MUSASHI of the Photon Factory, KEK [46–48]. The ultrathin films reported here were grown at a rate of 0.02 nm/s, as estimated from the thickness calibration of thick 10-nm VO$_2$ films using grazing incident x-ray reflectivity. The thicknesses of the films were regulated by deposition duration in the range of 0.5 to 10 nm. During the deposition, the substrate temperature was maintained at 400°C and the oxygen pressure was maintained at 10 mTorr. We remark here that we carefully optimized the growth temperature not only to avoid interdiffusion of the constituent transition metals across the interface but also to achieve the coherent growth of VO$_2$ films with high crystallinity. The surface structures and cleanness of the ultrathin VO$_2$ films were confirmed via reflection high-energy electron diffraction and core-level photoemission measurements, respectively. The results of the detailed characterization of the grown VO$_2$ films are presented in the Supplemental Material [49].

The surface morphologies of the measured films were analyzed via atomic force microscopy in air. The root-mean-square (RMS) values of the films' surface roughness $R_{RMS}$ were less than 0.2 nm (see Fig. S1 in Supplemental Material [49]), indicating that the thicknesses of these films were controlled to the scale of V-V dimer length (approximately 0.3 nm). The crystal structure was characterized by x-ray diffraction, which confirmed the coherent growth of the films on the substrates (see Fig. S2 in Supplemental Material [49]). The electrical resistivity was measured using the standard four-probe method.

PES measurements were performed *in situ* with the use of a VG-Scienta SES-2002 analyzer with total energy resolutions of 120 meV and 200 meV at photon energies of 700 eV and 1200 eV, respectively. The vacuum-transferring of the grown samples prevented the overoxidation of the surface layer (see Fig. S3 in Supplemental Material [49]) that had been an obstacle in previous studies on the thickness-dependence of VO$_2$ [36–42]. The XAS spectra were also measured *in situ* with linearly polarized light via measurement of the sample drain current. For linear dichroism measurements, we acquired the XAS spectra at angles between $\theta = 0°$ and 60° between the $c_R$ axis direction and the polarization vector while maintaining a fixed angle between the direction normal to the surface and the incident light [48]. $E_F$ of each sample was determined by measurement of a gold film that was electrically connected to the sample.

## III. RESULTS

### A. Formation of chemically abrupt interface

Before discussing the thickness dependence of the MIT, we present evidence for the fact that the prepared ultrathin $VO_2$ films had atomically and chemically abrupt $VO_2/TiO_2$ interfaces. This feature comprises a precondition to the present study on the confined films, and the evidence guarantees that the precondition is fulfilled. Figure 1 shows the Ti $2p$ core-level spectra of $VO_2/TiO_2$ taken at 320 K (corresponding to the rutile metallic phase of the thick 10-nm $VO_2$ film) with varying film thickness $t$ as well as a $TiO_2$ substrate as a reference. The spectra shown in Fig. 1(a) are normalized to the incident photon flux, reflecting the attenuation of the Ti $2p$ signal from buried $TiO_2$ by $VO_2$ overlayers. The intensity of the Ti $2p$ core level is steeply reduced with increasing $t$ and almost disappears at $t = 3$ nm, suggesting the formation of a chemically abrupt interface. The shift of energy position by 0.7 eV is attributed to the formation of a Schottky barrier at the heterointerface between rutile metallic $VO_2$ films and $n$-type oxide-semiconductor Nb:$TiO_2$ (see Fig. S4 in Supplemental Material [49]) [22,50–53]. In addition, the line shape of the Ti $2p$ core level maintains its original $Ti^{4+}$ feature, indicative of the invariance of the chemical environments even at the interface, although slight asymmetric spectral behavior is observed for thicker $VO_2$ films owing to the formation of the Schottky barrier (see Fig. S4 in Supplemental Material [49]) [54]. In the intermixed interface of a previous study [43], the significant broadening of the Ti $2p$ core level was observed owing to the formation of $Ti_xV_{1-x}O_2$ solid solution at the interface. Therefore, the observed invariance of the Ti $2p$ line shape strongly suggests the formation of a chemically abrupt interface in the present films.

To evaluate the length of the possible interdiffusion, we plot the Ti $2p$ core-level intensity $I_{Ti}$ as a function of the $VO_2$ overlayer thicknesses in Fig. 1(b) in comparison with a calculated photoemission attenuation function, $I_{Ti} = e^{-t/\lambda}$, where $t$ is the thickness of the $VO_2$ overlayer and $\lambda$ is the inelastic mean-free path of photoelectrons. The excellent agreement between the experiment and calculation indicates that the constituent cations are not intermixed with each other across the interface within the experimental margins. To highlight the chemical abruptness of the present films in comparison with the previous results, we simulated the intensity attenuation assuming the formation of an intermixing layer of 0.6 nm [39,40], 0.7 nm [41], and 3.1 nm [43]. The results are overlaid in the inset of Fig. 1(b). From the simulation results, the interdiffusion length of the present film is estimated to be less than 0.3 nm. Furthermore, the $R_{RMS}$ values of the films, as well as the substrate, were less than 0.2 nm irrespective of the film thickness (see Fig. S1 in Supplemental Material [49]). These results indicate the formation of an atomically and chemically abrupt interface in the present films as well as the flat surface essential for our spectroscopic measurements.

### B. Electronic phase diagram of ultrathin $VO_2$ films

As demonstrated, the present films have a chemically abrupt interface to a $TiO_2$ substrate and are grown coherently on the substrate with high crystallinities. Therefore, the next crucial issue is the intrinsic properties of the nanoscale $VO_2$ films. Figure 2 shows the temperature-dependent resistivity ($\rho$–$T$) curves for ultrathin $VO_2$ films grown on nondoped $TiO_2$ substrates with exactly the same growth conditions. For the thick 10-nm film, the typical behaviors across the MIT are observed; the $\rho$–$T$ curve steeply changes across the MIT accompanied by the thermal hysteresis characteristic of a first-order phase transition [27,37–41]. The MIT temperature $T_{MIT}$ is determined to be 293 K, while the change in resistivity across the MIT [$\rho$(250 K)/$\rho$(320 K)] is considerably larger than $10^3$. These values are almost the same as corresponding ones of previously reported epitaxial $VO_2$ films grown on $TiO_2$(001) substrates under in-plane tensile strain [27,37–41], which guarantees that the $VO_2$ films in the present work are of the same quality as those in previous studies. Furthermore, the steep change in the $\rho$–$T$ curve across the MIT suggests that the influence of the interdiffusion of the constituent transition metals across the interface is negligible in the present sample, as expected.

When film thickness $t$ is reduced, $T_{MIT}$ decreases to 287 K at $t$ = 3 nm from the original $T_{MIT}$ of 293 K for $t$ = 7–10 nm, and increases to 309 K beyond the original $T_{MIT}$ from around $t$ = 2 nm, accompanied by the reduction of the change in resistivity and the width of the hysteresis across the MIT. Even in the case of $t$ = 1.5 nm, the $\rho$–$T$ curve exhibits a broad MIT around 303 K with a concomitant small but detectable hysteresis loop. It should be noted that the behavior is consistent with that in a previous study on a $VO_2$/$TiO_2$ superlattice [44], suggesting the occurrence of crossover from collective to local V-V dimerization for $t$ = 1.5–2.0 nm. Eventually, the MIT itself seems to disappear for $t$ < 1.5 nm because the $\rho$–$T$ curves do not exhibit any kink structures, although the possibility for the occurrence of the MIT above the measurement temperature range is not completely eliminated. The results are summarized as the phase diagram in Fig. 2(b). The reentrant behavior of $T_{MIT}$ around few-nanometer thickness was also observed in previous thickness-dependent studies, although it has been considered as extrinsic effects, such as overoxidation and/or interdiffusion [36–43]. This unusual behavior may reflect the delicate balance between the confinement states of a band responsible for Peierls transition and the formation energy of V-V dimers, which we will discuss later in connection with the spectroscopic results.

**C. Electronic structure determined by photoemission spectroscopy**

Figure 3 shows the temperature dependence of the valence-band spectra of $VO_2$ ultrathin films grown on Nb:$TiO_2$ substrates by controlling the $VO_2$ layer thickness. Because the Nb:$TiO_2$ substrate, which is an *n*-type degenerated semiconductor, has a wide bandgap of 3.0 eV below $E_F$, as can be seen in Figs. 3(a) and 3(b), the electronic structures near $E_F$ of the ultrathin $VO_2$ films are not influenced by signals from the substrate, even for ultrathin films. These spectra mainly contain two features: structures derived from O 2*p* states at binding energies of 3–10 eV and peaks derived from the V 3*d* states near $E_F$ [12,55–57]. Regarding the thick $VO_2$ films ($t$ = 10 nm), the spectra exhibit the characteristic features representative of the MIT of $VO_2$ [12]; the spectrum near $E_F$ in the high-temperature (HT) metallic phase (measured at $T$ = 320 K) consists of a sharp coherent peak just at $E_F$ and a weak broad satellite structure around 1.2 eV, while that in the low-temperature (LT) insulating phase (at $T$ = 250 K) shows a single peak around 1.0 eV, which leads to the formation of an energy gap at $E_F$. Furthermore, focusing on the O 2*p* states, we observe dramatic changes across the MIT. These changes are responsible for the structural changes concomitant with the MIT in $VO_2$ [12].

Even when the thickness is reduced down to 2 nm, the line shapes of these spectra are almost identical to each other; the characteristic features representative of the MIT of $VO_2$ remain, whereas there is a slight reduction in the intensity just at $E_F$ for the HT phase. The spectral behaviors correspond to the fact that the conductivity gradually reduces with decreasing $t$, although the fundamental behaviors of the MIT are maintained in the thickness-dependent $\rho$–$T$ curves [see Fig. 2(a)]. When the film thickness further decreases below 2 nm, the spectra exhibit remarkable and systematic changes mainly in V 3*d* states near $E_F$. To investigate the spectral changes near $E_F$ in more detail, we present the spectra near $E_F$ according to the enlarged binding-energy scale in Fig. 3(b). For $t = 1.0$–$1.5$ nm, the intensity of the V $3d$ derived coherent peak at $E_F$ for the HT phase is considerably reduced. In addition, the leading edge of the V $3d$ states appears to shift from above $E_F$ to below $E_F$ at $t = 1.5$ nm, suggesting the evolution of a pseudogap at $E_F$ [33,34]. With further decreasing $t$, the spectral weight at $E_F$ becomes negligible for $t = 1.0$ nm, and eventually a clear energy gap opens at $t = 0.5$ nm. An extrapolation of the linear portion of the leading edge to the energy axis yields a valence-band maximum of 450 meV for $t = 0.5$ nm and almost 0 meV for $t = 1.0$ nm. The negligibly small residual spectral weight at $E_F$ for $t = 1.0$ nm may be due to the finite energy resolution of our experimental system (120 meV). These results indicate the occurrence of thickness-dependent MIT at a critical thickness of 1.0–1.5 nm for the HT phase.

Accompanying the thickness-dependent MIT for the HT phase, the spectral changes across the temperature-dependent MIT also weaken, as can be seen in Fig. 3(a). Interestingly, the spectral change in V $3d$ states near $E_F$ is still visible even at $t = 1.0$ nm, irrespective of the disappearance of the metallic states, implying the existence of local V-V dimerization in this thickness region [44]. Meanwhile, the thickness dependence of the LT phase is different from that of the HT phase. The spectra at the LT phase maintain their shape even at $t = 1.5$ nm, whereas the pseudogap is formed at the HT phase. For $t \leq 1.0$ nm, there are no significant changes in the LT spectra at first glance, except the lowering of their intensity due to a reduction of the relative number of V ions. However, a closer look reveals that the peak position shifts from approximately 1.0 eV for $t = 1.5$–$10$ nm to 1.3 eV for 0.5–1.0 nm, suggesting that the LT monoclinic insulating phase also changes into another insulating phase emerging in the thinner region.

### D. V-V dimerization studied by oxygen $K$-edge x-ray absorption spectroscopy

Focusing on the O $2p$ states in the energy range of 3.0–10 eV, dramatic changes across the MIT are observed down to $t = 1.5$ nm, but they considerably weaken at $t = 1.0$ nm, and eventually changes other than the thermal broadening are barely visible at $t = 0.5$ nm. The crossover of some kind across $t = 1.0$ nm seems to be related to the peak-position shift in states near $E_F$ at the LT phase [Fig. 3(b)]. Because the changes in the O $2p$ states are responsible for the structural change accompanied by the MIT in $VO_2$ [12], these results strongly suggest that such a structural change disappears, and another insulating phase emerges in this ultrathin thickness range. To elucidate the crossover of the structural transition (V-V dimerization), we have measured the polarization dependence of oxygen $K$-edge XAS, which has previously been utilized as a good indicator of V-V dimerization [12,39,42,47,48,58–60], as shown in Fig. 4(a). The XAS at the O $K$ edge is a technique complementary to PES for investigating the electronic structures in conduction bands via probing of the unoccupied O $2p$ partial density of states that are mixed with the unoccupied V $3d$ states. Because the V-V dimerization in the monoclinic phase splits a half-filled $d_{//}$ state into occupied $d_{//}$ and unoccupied $d_{//}^*$ states [3], an additional peak corresponding to the $d_{//}^*$ state appears in the XAS spectra only for the insulating monoclinic phase, as schematically illustrated in Fig. 4(b). Furthermore, owing to the strict dipole selection rule, the additional $d_{//}^*$ states only appear in the spectra acquired with the polarization vector $E$ parallel to the $c_R$ axis ($E$ $//$ $c_R$). Indeed, as can be confirmed from the $t = 10$ nm spectra shown in Fig. 4(a), the $d_{//}^*$ peak emerges at 530.6 eV (indicated by the solid triangle) in the insulating monoclinic phase ($T = 250$ K) measured at the $E$ $//$ $c_R$ geometry, whereas it disappears in the metallic rutile one ($T = 320$ K). Furthermore, the identification of the $d_{//}^*$ states is confirmed by inferring the polarization dependence (i.e., linear dichroism) of the XAS spectra; the additional $d_{//}^*$ peak in the LT phase disappears for the spectrum taken with $E$ perpendicular to the $c_R$ axis ($E \perp c_R$). Thus, the existence of the $d_{//}^*$ peak in the spectra at the $E$ $//$ $c_R$ geometry can be used as a fingerprint of the V-V dimerization in $VO_2$.

As can be seen in Fig. 4(c), with decreasing $t$, the $d_{//}^*$ states appearing at the LT monoclinic insulating phase begin to weaken around $t = 2$ nm and almost disappear at $t \leq 1.0$ nm. The thickness dependence is consistent with that of the O $2p$ states in valence-band structures [Fig. 3(a)]. Therefore, these spectroscopic results indicate that V-V dimerization no longer collectively occurs for ultrathin $VO_2$ films with $t \leq 1.0$ nm; in other words, the ultrathin films would be in the rutile insulating phase. The emergence of a new phase associated with the disappearance of the collective V-V dimerization is also confirmed by the leading-edge shift at the absorption edge. As can be seen in Fig. 4(c), the leading-edge shift is clearly observed across the temperature-dependent MIT for $t = 1.5$–10 nm at the photon energy of 529.0 eV, which is the counterpart of the energy-gap formation in the PES results in Fig. 3. The temperature-induced edge shift seems to disappear for $t \leq 1.0$ nm, and the energy position of the edge itself slightly shifts to a higher photon-energy side by 0.3 eV. The thickness-induced edge shift is consistent with the peak position shift observed in thickness-dependent measurements for the LT phase [Fig.

3(b)]. In connection with the PES results, it is naturally concluded that the LT monoclinic insulating phase changes into another insulating phase without V-V dimerization in the thinner region. The emergence of the new phase in the thin limit has an important implication for the thickness-dependent electronic and structural changes in $VO_2$ ultrathin films.

## IV. DISCUSSION

Finally, we discuss the roles of Mott and Peierls instabilities in the observed thickness-dependent MIT. The thickness-dependent physical properties of ultrathin $VO_2$ films should be responsible for the delicate balance between the Mott and Peierls instabilities. As the film thickness decreases, there are two effects that may diminish the cooperative Mott–Peierls (or Peierls–Mott) transition: the increment of electron–electron correlation (i.e., Mott instability) due to dimensional crossover [33–35] and the reduction of a bandlike Peierls instability due to quantization of the $d_{//}$ band [29–32]. As for a Mott instability due to the dimensional crossover effect, it is a kind of a bandwidth-controlled Mott transition; a decrease in the layer thickness of ultrathin films causes a reduction in the effective coordination number of constituent ions at the interface and surface. The resultant reduction of the effective band width $W$ from a three-dimensional thick film to the two-dimensional ultrathin film drives MIT in conductive transition metal oxides, while the on-site Coulomb repulsion $U$ does not change. In fact, the behavior of the thickness-dependent MIT observed for the HT metallic phase appears ubiquitous for conductive transition metal oxides [33,34]: The pseudogap is formed at a scale of a few nanometers and eventually evolves to an energy gap at $E_F$ as a Mott gap. Thus, in the present case, the Mott ground state ($U/W \ll 1$) may be stabilized below the critical film thickness of 1.0–1.5 nm, and $VO_2$ become a rutile insulator with a Mott gap in the thin limit.

The other effect is the confinement effect of the $d_{//}$ band that has one-dimensional character along the $c_R$ direction. The energy gain from the bandlike Peierls transition should be reduced by the quantization of the $d_{//}$ band, resulting in the suppression of Peierls-assisted phenomena. As can be seen in the spectroscopic results shown in Figs. 3 and 4, the characteristic spectral changes associated with the bandlike Peierls phenomena [12] may disappear for $t \leq 1.0$ nm, although the weak spectral change probably due to local V-V dimerization still exists for $t = 1.0$ nm. These results suggest that there is insufficient energy gain from a bandlike Peierls effect to induce collective structural phase transition for $t \leq 1.0$ nm. In contrast, the increment of electron–electron correlation ($U/W$) due to dimensional crossover further stabilizes the Mott ground states. As a result of the superiority of Mott instability over the Peierls one, the VO$_2$ films become a Mott insulator with rutile structure in the thin limit of $t \leq 1.0$ nm.

The delicate balance between Mott instability and bandlike Peierls instability is modulated at a scale of a few nanometers by the dimensional crossover effects, which may consequently induce the complicated electronic phase diagram of the ultrathin VO$_2$ films. The present thickness-dependent studies may offer valuable insight into the longstanding issue of determining the precise relationship between Mott and Peierls instabilities in VO$_2$. In addition, from an application perspective, the determination of the critical thickness in VO$_2$ provides important information for designing the VO$_2$ channel layer of the Mott transistor with a thickness comparable to the electrostatic screening length from the interface (Thomas-Fermi screening length from a conventional solid-state gate dielectric) [22,61,62]. To gain a more comprehensive understanding of the origins of thickness-dependent behavior at the few-nanometer scale, further systematic investigation is required. In particular, investigation of different crystallographic orientations [63,64] will be necessary to examine the effects of two instabilities on the MIT and the resultant complicated phase diagram.

## V. CONCLUSION

We determined the electronic phase diagram of ultrathin VO$_2$ films coherently grown onto TiO$_2$(001) substrates. The formation of a chemically abrupt interface, which is a key issue to study the intrinsic physical properties of nanostructured VO$_2$ films, has been confirmed by core-level measurements. Subsequently, we investigated the change in electronic and crystal structure (V-V dimerization) of dimensionality-controlled VO$_2$ films via *in situ* PES and XAS measurements. The spectra exhibited remarkable and systematic changes as a function of VO$_2$ film thickness, which is in line with the transport properties: (1) Collaborative Mott–Peierls MIT characteristic to bulk VO$_2$ almost holds down to 2.0 nm, although there is modulation of $T_{MIT}$ accompanied by the reduction of the change in resistivity and the width of the hysteresis across the MIT. (2) In the case of films with thickness of 1.5–2.0 nm, a pseudogap evolves at $E_F$ for the HT metallic phase, and the spectral change across the MIT consequently weakens but still exists, being concomitant with the collective V-V dimerization. (3) The pseudogap finally evolves into an energy gap when the film thickness is reduced to less than 1.5 nm, indicating the occurrence of thickness-dependent MIT at the critical thickness of 1.0–1.5 nm. Furthermore, $VO_2$ films with thicknesses of 1.0 nm or less exhibit a rutile-insulating nature without V-V dimerization. These results demonstrate that the collaborative Mott–Peierls (or Peierls–Mott) transition persists down to 1.5 nm, whereas the insulating states in the thin limit may be better described in terms of Mott insulating states.

## ACKNOWLEDGEMENTS

The authors are very grateful to Y. Kuramoto for our helpful discussions. This work was financially supported by a Grant-in-Aid for Scientific Research (No. 16H02115 and No. 16KK0107) from the Japan Society for the Promotion of Science (JSPS), CREST (JPMJCR18T1) from the Japan Science and Technology Agency (JST), and the MEXT Element Strategy Initiative to Form Core Research Center (JPMXP0112101001). The preliminary sample characterization using hard x-ray photoemission at SPring-8 was conducted under approval of the Japan Synchrotron Radiation Research Institute (2019B1248). The work performed at KEK-PF was approved by the Program Advisory Committee (proposals 2019T004 and 2018S2-004) at the Institute of Materials Structure Science, KEK.

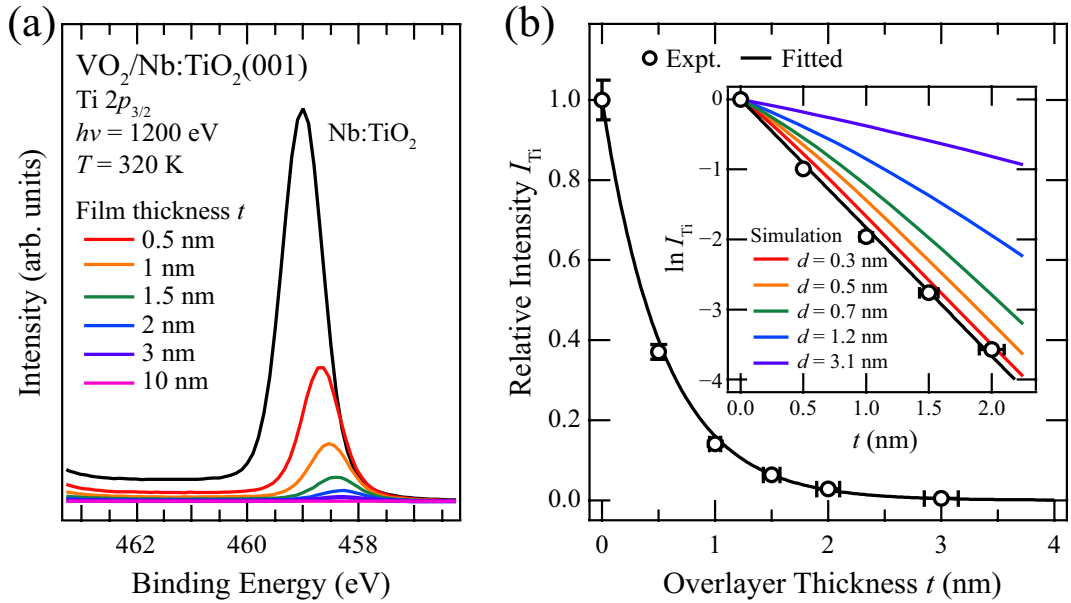

**FIG. 1.** (Color online) (a) Thickness dependence of Ti $2p_{3/2}$ core-level spectra measured at $h\nu =$ 1200 eV for $VO_2$/Nb:$TiO_2$(001) ultrathin films. Each spectrum is normalized to the incident photon flux; hence, the intensity reduction with increasing $VO_2$ overlayer thickness $t$ reflects the attenuation of the Ti $2p$ signal from buried $TiO_2$ by the overlayer. (b) Plot of relative intensities of the background-subtracted Ti $2p_{3/2}$ core-level spectra, $I_{Ti}$, as a function of $VO_2$ overlayer thickness. The experimental points are fitted to the photoemission attenuation function with the photoelectron inelastic mean-free path of 0.55(2) nm. The inset represents the logarithm plot of $I_{Ti}$ with respect to $t$ in comparison with the simulation curves assuming the intermixing of V and Ti ions at the interface with interdiffusion lengths $d$ of 0.3, 0.5, 0.7 [41], 1.2, and 3.1 nm [43]. From the comparison, it is clear that the interdiffusion length of the present film is less than 0.3 nm, which is much smaller than that of previous reports and comparable to the V-V dimer length of approximately 0.3 nm.

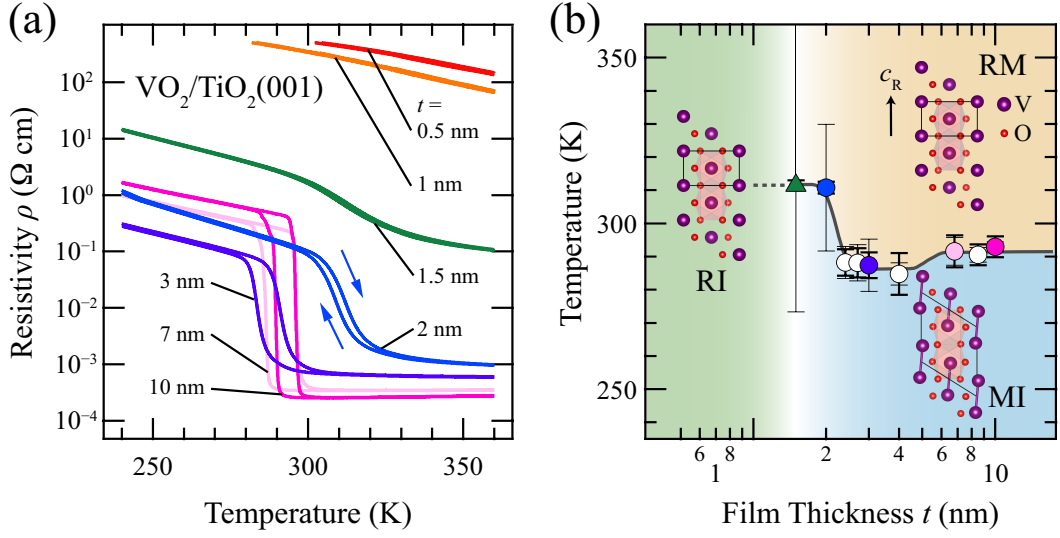

**FIG. 2.** (Color online) (a) Temperature dependence of resistivity ($\rho$–$T$) for VO$_2$/TiO$_2$(001) ultrathin films with thicknesses $t$ varying from 0.5 to 10 nm. Note that the saturated behavior of the resistivity value at lower temperatures for $t$ = 0.5 and 1.0 nm is due to the limitation of our apparatus. (b) Electronic phase diagram of VO$_2$(001) ultrathin films as functions of temperature and $t$. The filled and open circles indicate $T_{\text{MIT}}$ determined from the $\rho$–$T$ curves. Here, $T_{\text{MIT}}$ was defined as the center of the hysteresis loop, namely, an average of the two inflection points in $\rho$–$T$ curves during cooling and heating. The thick and thin bars in the vertical axis represent the hysteresis and transition width in the $\rho$–$T$ curves [40], respectively. Note that the colors of each filled symbol correspond to those in the $\rho$–$T$ curves. Solid and dotted lines are merely guides for ease of visualization. The monoclinic insulating, rutile metallic, and rutile insulating phases are denoted by MI, RM, and RI, respectively, which are assigned from the spectroscopic results discussed later. The insets show the crystal structures of corresponding rutile and monoclinic VO$_2$.

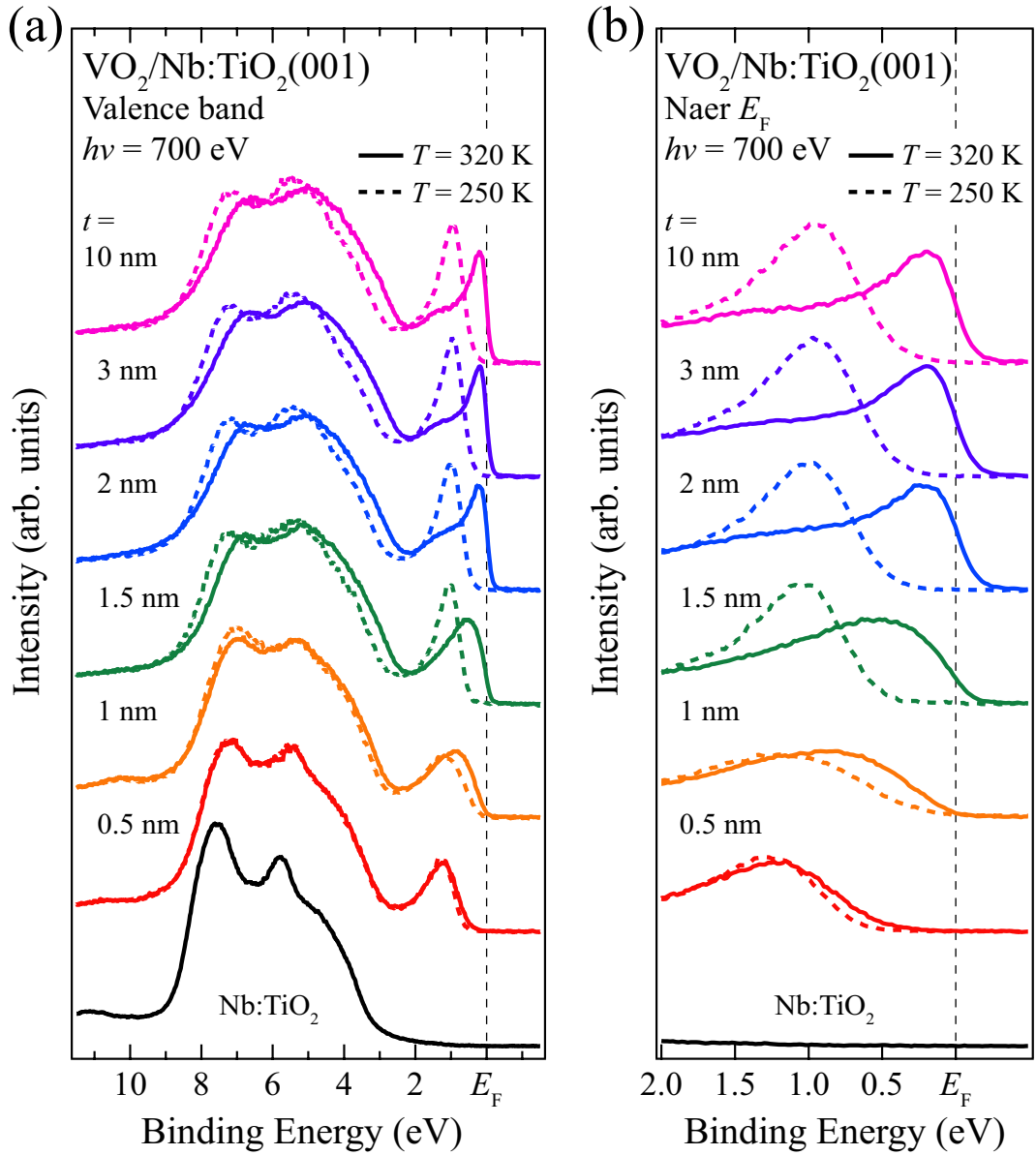

**FIG. 3.** (Color online) (a) Thickness dependence of valence-band spectra measured at $T = 320$ K (HT rutile metallic phase for $t = 10$ nm) and 250 K (LT monoclinic insulating phase for $t = 10$ nm) for the $VO_2/Nb:TiO_2(001)$ ultrathin films. (b) PES spectra near $E_F$ in an expanded energy scale. Note that the colors of each spectrum correspond to those in Figs. 1 and 2.

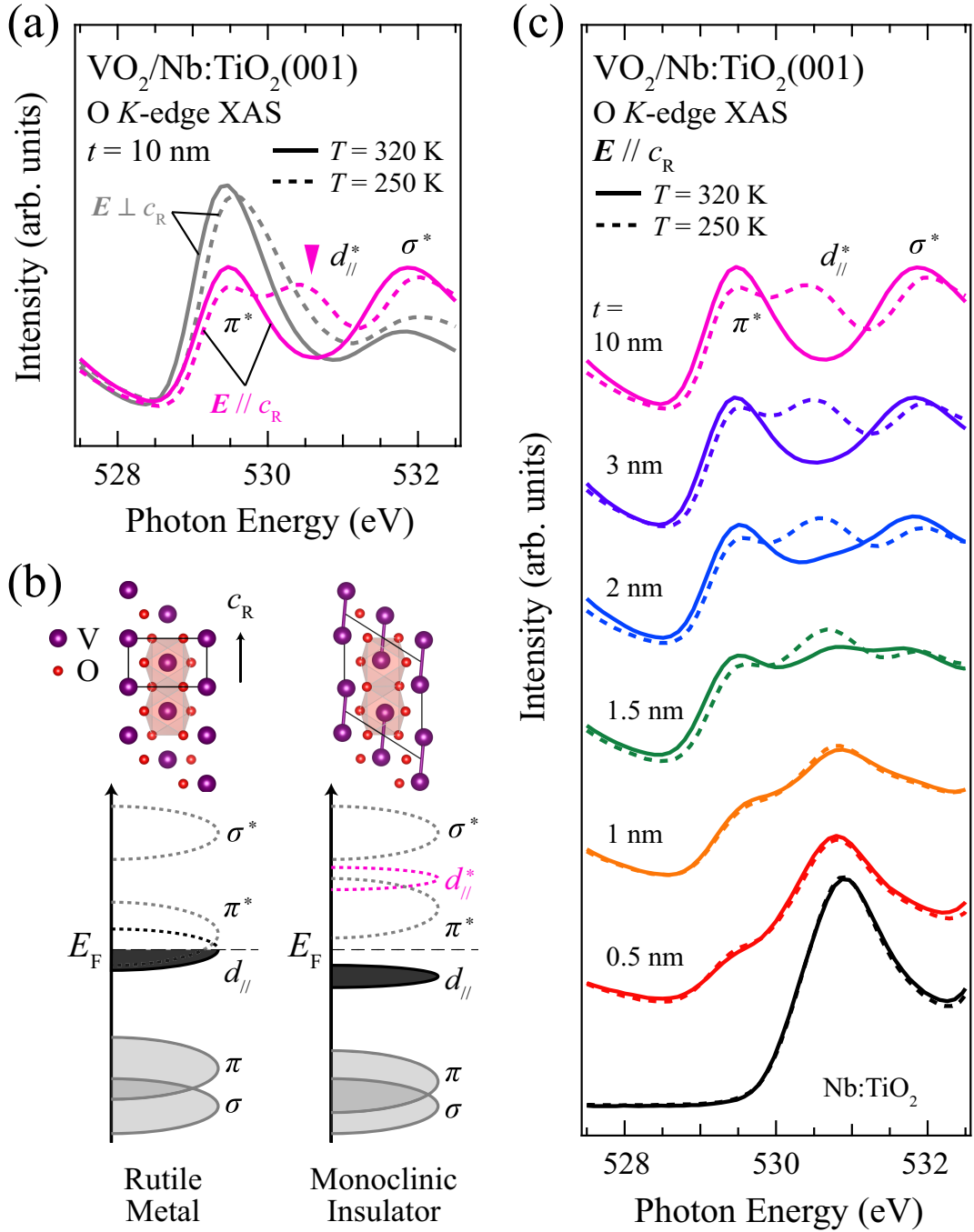

**FIG. 4.** (Color online) (a) Temperature dependence of O $K$-edge XAS spectra with different polarizations of $VO_2$/Nb:$TiO_2$(001) films with $t$ = 10 nm. XAS spectra acquired with the polarization vector $\boldsymbol{E}$ perpendicular to the $c_R$ axis ($\boldsymbol{E} \perp c_R$) and parallel to the $c_R$ axis ($\boldsymbol{E} // c_R$) are represented by gray and magenta lines, respectively. The XAS spectra with $\boldsymbol{E} // c_R$ ($I_{//}$) are deduced from the expression $I_{//} = (4/3)(I - I_{\perp}/4)$, where $I_{\perp}$ (namely, that corresponding to $\boldsymbol{E} \perp$

$c_R$) and $I$ denote the spectra measured with normal ($\theta = 0°$) and grazing ($\theta = 60°$) incidence, respectively [48]. Following the assignments made in previous studies [58], those are schematically illustrated in (b) together with the corresponding crystal structures, the first peak around 529.5 eV can be assigned to $\pi^*$ bands formed by V $3d_{xz}$ and $3d_{yz}$ orbitals, while the second peak can be assigned to $\sigma^*$ bands formed by $3d_{3z^2-r^2}$ and $3d_{x^2-y^2}$ orbitals. The additional peak that emerges around 530.6 eV only for the spectra of insulating monoclinic phase ($T = 250$ K) with $\boldsymbol{E} \mathbin{/\mkern-4mu/} c_R$ can be assigned to the $d_{//}^*$ state due to V-V dimerization (indicated by filled triangle). (c) Thickness dependence of O $K$-edge XAS spectra acquired with $\boldsymbol{E} \mathbin{/\mkern-4mu/} c_R$ at $T = 320$ and 250 K for the VO$_2$/Nb:TiO$_2$(001) ultrathin films. Note that the colors of each spectrum correspond to those in Figs. 1–3.

# Supplemental Material

# Thickness dependence of electronic and crystal structures in $VO_2$ ultrathin films: suppression of the collaborative Mott–Peierls transition

D. Shiga[1,2], B. E. Yang[1], N. Hasegawa[1], T. Kanda[1], R. Tokunaga[1], K. Yoshimatsu[1], R. Yukawa[2], M. Kitamura[2], K. Horiba[2], and H. Kumigashira[1,2,*]

[1] *Institute of Multidisciplinary Research for Advanced Materials (IMRAM), Tohoku University, Sendai, 980–8577, Japan*

[2] *Photon Factory, Institute of Materials Structure Science, High Energy Accelerator Research Organization (KEK), Tsukuba, 305–0801, Japan*

[*]Author to whom correspondence should be addressed: kumigashira@tohoku.ac.jp

## I. Sample characterization

### A. Surface morphology

The preconditions for the thickness-dependent study of $VO_2$ films are atomically flat surfaces and chemically abrupt $VO_2/TiO_2$ interfaces. The atomically flat surfaces of all the measured samples were confirmed by *ex situ* atomic force microscopy (AFM). The AFM images are shown in Fig. S1. As can be seen, the AFM images show similar surface morphologies irrespective of film thickness. The values of the root-mean-square (RMS) of the surface roughness $R_{RMS}$ estimated from the AFM images were all less than 0.2 nm. The values are almost the same as that of the original $TiO_2$ substrate ($R_{RMS}$ = 0.13 nm). The $R_{RMS}$ values of the measured ultrathin films were all less than V-V dimer length (approximately 0.3 nm), indicating that the thicknesses of these films were controlled to the scale of V-V dimer length and that the smooth surface and interface were maintained as the film was grown to 10 nm.

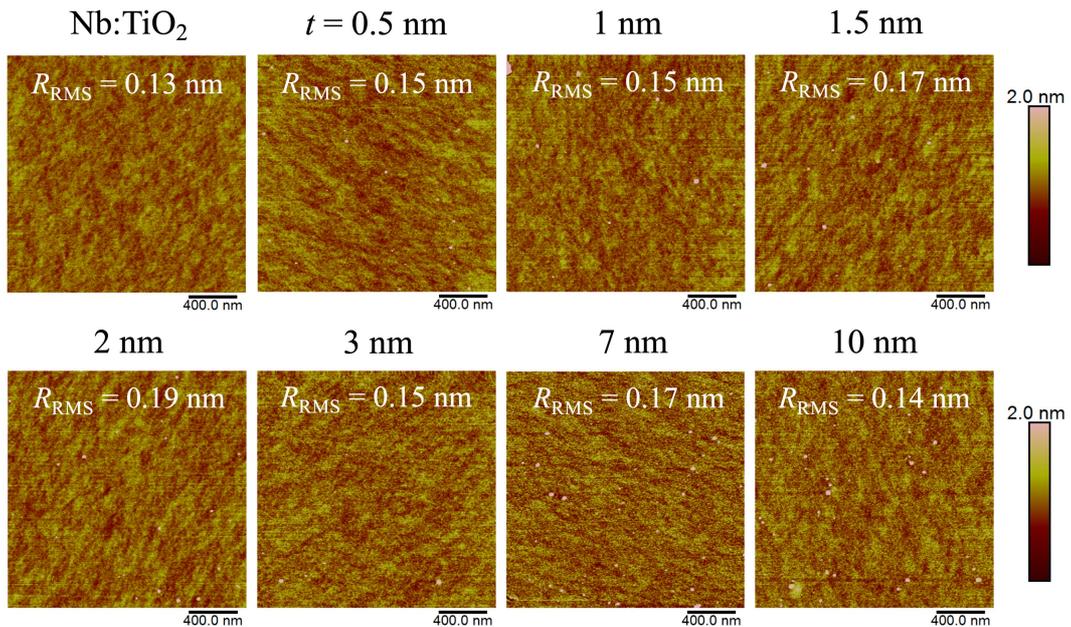

**Fig. S1.** Typical AFM images of a $TiO_2(001)$ substrate and $VO_2(001)$ ultrathin films with different thickness $t$ after growth.

## B. Crystal structure

Crystal structures of VO$_2$ ultrathin films were characterized by x-ray diffraction (XRD) measurements. Figure S2(a) shows typical XRD patterns for ultrathin films. For $t$ = 10 nm, the XRD pattern with well-defined Laue fringes is clearly observed around the 002 diffraction peak, indicating the formation of abrupt interfaces as well as the high quality of the VO$_2$/TiO$_2$ heterostructures. In contrast, the Laue fringes are barely visible for $t \leq 3$ nm because the thicknesses are too thin to fulfill the reflection conditions, as demonstrated by the simulations. The high crystallinity of VO$_2$ films is also confirmed by the rocking curve of the 002 diffraction. As shown in Fig. S2(b), the full-width at half-maximum (FWHM) of the rocking curve for the 002 diffraction is evaluated to be around 0.038°, which is almost the same as that previously reported for a VO$_2$ film with high crystallinity and coherently grown on TiO$_2$(001) substrates [S1,S2]. Besides, the coherent growth of VO$_2$ ultrathin films on TiO$_2$(001) substrates was confirmed by reciprocal space mapping around the 112 reciprocal point, as shown in Fig. S2(c). The out-of-plane ($c_R$ axis) lattice constant of rutile VO$_2$ thin films pseudomorphically grown on the isostructural TiO$_2$(001) substrate is 0.2834(5) nm, while the in-plane lattice constant is 0.4593(2) nm. These cryastallographic results identify the current PLD-grown films as being highly crystalline.

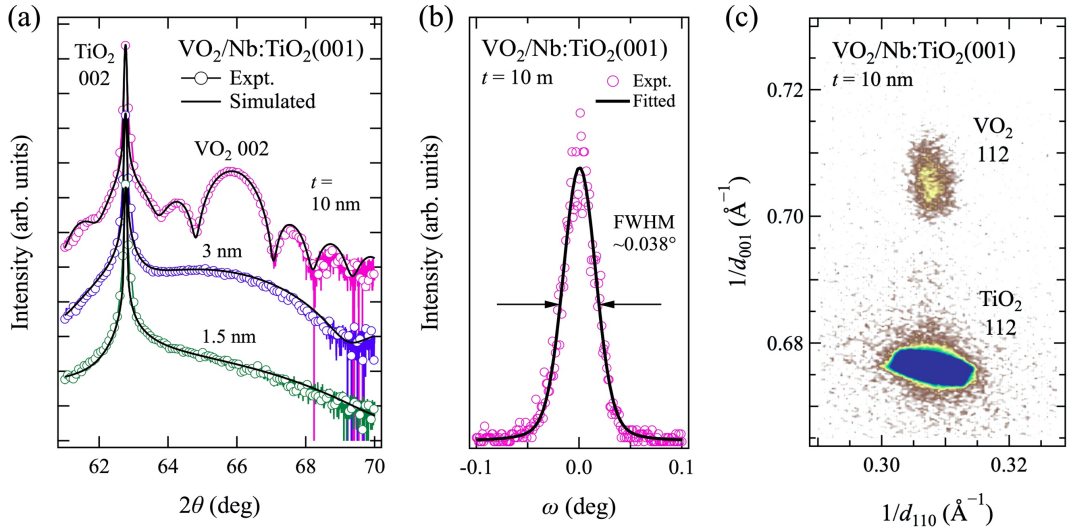

**Fig. S2.** (a) Typical XRD patterns for VO$_2$ ultrathin films on TiO$_2$(001) substrates with $t$ = 1.5, 3, and 10 nm. (b) Corresponding rocking curve of the 10-nm VO$_2$ film. (c) Reciprocal space mapping around the 112 reciprocal point of the 10-nm VO$_2$ film.

## II. V 2p core-level spectra

Figure S3(a) shows V 2p core-level spectra of the 10-nm $VO_2$ film transferred from the deposition chamber to the analysis chamber in an ultrahigh vacuum (*in situ*) and after exposure to air (*ex situ*). The *ex situ* spectrum is taken at low-temperature (LT) monoclinic insulating phase (250 K), while the *in situ* spectra are taken both at high-temperature (HT) rutile metallic phase (320 K) and LT monoclinic insulating phase. For the *in situ* spectra, the characteristic changes are observed across the temperature-dependent metal–insulator transition (MIT), reflecting the structural phase transition from the HT rutile to LT monoclinic phases [S3]; the broad structure around 515.6 eV with a shoulder structure at the lower binding energy side in the rutile metallic phase becomes a sharp single peak in the monoclinic insulating phase. According to the cluster model calculation [S3], the shoulder structure around 514.2 eV is assigned to well-screened features that originate from core-hole screening by a coherent band (metallic states) at $E_F$ [S4].

On the other hand, for the *ex situ* spectrum, the $V^{5+}$ states emerge at 517.2 eV owing to the overoxidation of the surface. The $V^{5+}$ states are barely visible in the *ex situ* hard x-ray photoemission spectra (HAXPES), as shown in Fig. S3(b). Furthermore, the temperature dependence of HAXPES spectra is almost the same as that of the *in situ* spectra. These results indicate that the overoxidation to $V^{5+}$ states occurs only at the surface of $VO_2$ films.

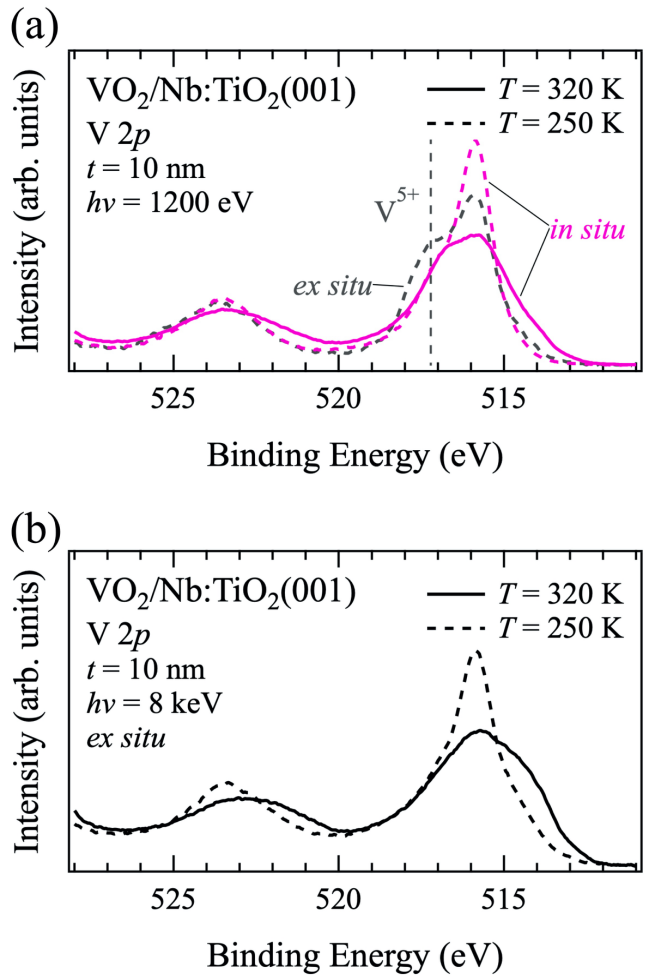

**Fig. S3.** (a) Comparison of V 2p core-level spectra for *in situ* transferred $VO_2$ films and *ex situ* ones. The *ex situ* spectrum is taken at LT monoclinic insulating phase, while the *in situ* spectra are taken both at HT rutile metallic phase and LT monoclinic insulating phases to confirm the characteristic change associated with the MIT. The energy position of $V^{5+}$ states due to the surface overoxidation in *ex situ* spectra is indicated by the vertical dotted line. (b) V 2p core-level spectra measured by HAXPES for *ex situ* transferred $VO_2$ films.

## III. Ti 2p core-level spectra

Figure S4(a) shows the Ti 2$p$ core-level spectra of VO$_2$/TiO$_2$ taken at 320 K (corresponding to the rutile metallic phase of thick VO$_2$ films) with varying film thickness $t$ as well as a TiO$_2$ substrate as a reference. The spectra are identical to those shown in Fig. 1(a) but normalized to peak intensity so that the peak shift due to band bending can be visualized. The peak shifts to the higher binding energy side from $t = 0$ nm (TiO$_2$ substrate) with increasing $t$ and almost saturates at $t = 2$–3 nm owing to the formation of a Schottky barrier at the heterointerface between rutile metallic VO$_2$ films and $n$-type oxide-semiconductor Nb:TiO$_2$. Judging from the saturation level of the peak shift, the energy shift due to band bending is estimated to be 0.7 eV, which is in good agreement with previous reports [S5,S6,S7].

In Fig. S4(b), we align the peak position of the peak-normalized spectra in Fig. S4(a) to visualize the change in line shape. The line shape of Ti 2$p$ core level maintains its original Ti$^{4+}$ feature, indicative of the invariance of the chemical environments even at the interface, although slight asymmetric spectral behavior is observed for thicker VO$_2$ films owing to the formed Schottky barrier potential. The invariance of chemical environment of the buried interface further supports the formation of the chemically abrupt interface in the VO$_2$/TiO$_2$ films.

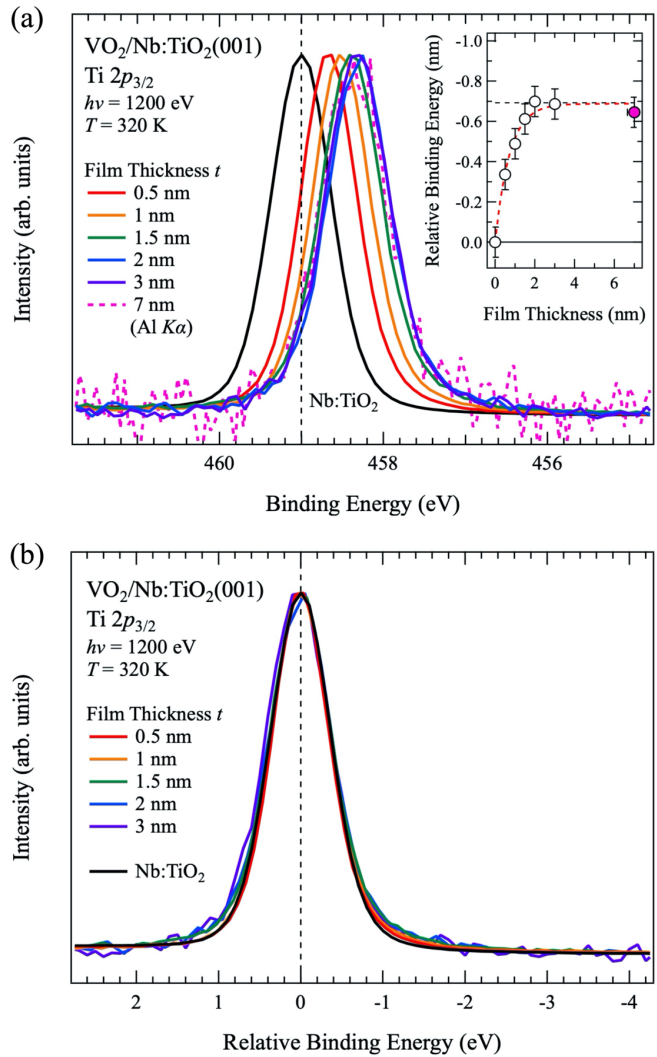

**Fig. S4.** (a) Ti 2$p$ core-level spectra of VO$_2$/Nb:TiO$_2$ with varying VO$_2$ overlayer thickness $t$ as well as a TiO$_2$ substrate as a reference. The inset shows a plot of the energy shift of the Ti-2$p$ peaks as a function of $t$, wherein only a magenta-colored data marker for $t = 7$ nm is acquired via *in situ* x-ray PES ($h\nu = 1486.6$ eV). The red dotted line is a guide for ease of visualization. (b) Ti 2$p$ core-level spectra aligned by the peak position.